\documentstyle[12pt]{article}
\renewenvironment{thebibliography}[1]
{\normalsize
 \begin{list}{[\arabic{enumi}]}
 {\usecounter{enumi} \setlength{\parsep}{0pt}
  \setlength{\itemsep}{3pt} \settowidth{\labelwidth}{[#1]}
  \sloppy}}
{\end{list}}

\newcommand{\cleqn}{\setcounter{equation}{0}}

\newcommand{\pr}{\hspace{\parindent}}
\newcommand{\simgt}{\hbox{\raise3pt\hbox to 0pt{$>$}\raise-3pt\hbox{$\sim$}}}
\newcommand{\simlt}{\hbox{\raise3pt\hbox to 0pt{$<$}\raise-3pt\hbox{$\sim$}}}
\def\lsa{\rlap{\lower 3.5 pt \hbox{$\mathchar \sim$}} \raise 1pt \hbox {$<$}}
\def\rsa{\rlap{\lower 3.5 pt \hbox{$\mathchar \sim$}} \raise 1pt \hbox {$>$}}

\begin{document}

\hfill\vbox{\baselineskip14pt
            \hbox{\bf KEK-TH-419}
            \hbox{\bf KEK Preprint 94-158}
            \hbox{November 1994}
            \hbox{\bf H}}
\baselineskip18pt

\vspace*{1cm}
\begin{center}
{\Large{\bf CP Violation in $\tau\rightarrow 3\pi\nu_\tau$}}
\end{center}
\vspace*{1cm}

\begin{center}
{\large S.Y.~Choi, K. Hagiwara, M.~Tanabashi}
\end{center}

\begin{center}
{\it Theory Group, KEK, Tsukuba, Ibaraki 305, Japan }
\end{center}
\vspace*{1cm}

\begin{center}
\large Abstract
\end{center}
\vspace{0.5cm}
\begin{center}
\begin{minipage}{14cm}
\baselineskip=18pt
\noindent
We consider CP violating effects in the decays
$\tau\rightarrow (3\pi)\nu_\tau$ where both the
${\rm J}^{\rm P}=1^+$ resonance, $a_1$, and ${\rm J}^{\rm P}=0^-$
resonance, $\pi^\prime$, can contribute.
The interference between the $a_1$ and $\pi^\prime$ resonances
can lead to enhanced CP-violating asymmetries whose magnitudes
depend crucially on the $\pi^\prime$ decay
constant, $f_{\pi^\prime}$.
We make an estimate of  $f_{\pi^\prime}$ with a simplified
chiral Lagrangian coupled to a massive pseudoscalar field, and
we compare the estimates from  the non-relativistic quark model
and from the QCD sum rule with the estimate from the `mock' meson model.
We then estimate quantitatively the size of CP-violating effects in a
multi-Higgs-doublet model and scalar-leptoquark models.
We find that, while CP-violating effects in the scalar-leptoquark
models may require more than $10^{10}$ $\tau$ leptons,
CP-violating effects from the multi-Higgs-doublet
model can be seen at the $2\sigma$ level with about $10^7$ $\tau$
leptons using the chiral Lagrangian estimate of
$f_{\pi^\prime}=(1\sim 5)\times 10^{-3}$ GeV.
\end{minipage}
\end{center}
\vfill

\baselineskip=18pt
\normalsize

\newpage
\setcounter{page}{2}

\section{Introduction}
\cleqn

\pr
The $\tau$ has the same interaction structure as the $e$
and the $\mu$ in the Standard Model (SM), apart from their masses.
However, for practical purposes\cite{Gomez}
the $\tau$ lepton, the most massive of the known leptons, behaves quite
differently from the $e$ and $\mu$ leptons in that (i)
the $\tau$ has hadronic decay modes\cite{Tsai,Pi}
(e.g. $\tau\rightarrow \pi\nu, \rho\nu, a_1\nu,...$) which
allow an efficient measurement of its polarization\cite{Tsai,Pi,Jadach}
and (ii) the couplings of the $\tau$ to neutral and charged Higgs
bosons\cite{Soni2,Grossman,Falk} are expected to dominate those
of the $e$ and $\mu$.
These features make the $\tau$ a rather special experimental
probe of new physics.

One phenomenon where new physics can play a crucial role is CP
violation. Currently CP violation has been detected only in the
$K$ meson system. The SM explains the effect adequately
through a phase in the Kobayashi-Maskawa (KM) matrix\cite{Kobayashi}.
However, it is also possible that other sources of CP violation
exist in nature.
Recently the $\tau$ decays into hadrons have been considered as probes
of such a non-KM-type of CP violation in the scalar sector of physics
beyond the SM.
Nelson {\it et al.}\cite{Nelson} have considered the so-called stage-two
spin-correlation functions to detect CP violation in the decay
$\tau\rightarrow (2\pi)\nu_\tau$, while in the same two-pion decay mode
Tsai\cite{Tsai1} has suggested tests of CP violation
with longitudinally polarized electron and positron
beams at $\tau$-Charm factories.
On the other hand, Kilan {\it et al.}\cite{Kilan} have studied
$\tilde{\rm T}$-odd triple momentum correlations in $\tau$ decays into
$K\pi\pi$ and $K\pi K$.

In the present work we consider the possibility of probing CP violation
in the decay of the $\tau$ into three charged pions,
$\tau\rightarrow (3\pi)\nu_\tau$, where both the $J^P=1^+$ resonance,
$a_1$, and $J^P=0^-$ resonance, $\pi^\prime$, can contribute.
In particular, we investigate whether the large widths of these
resonances in the decay of the $\tau$ can be used to enhance
CP-violation effects in extensions of the SM
with scalar-fermion interactions which are consistent with the symmetries
of the SM; the importance of broad resonances has been emphasized
in the context of the top quark\cite{Top} and the
$B$ meson\cite{Bmeson} in the last few years.

In order to observe CP-violating effects there should exist not only
a CP-violating phase but also processes interfering with
different CP-conserving  phases. In $\tau$ decays one can in general have
a CP-violating phase between the $W$-exchange diagram and scalar-exchange
diagrams in extensions of the SM such as a multi-Higgs-doublet
models and  scalar-leptoquark models.
On the other hand, resonance enhancements of transition amplitudes and their
coherent superposition  can provide a large CP-conserving phase
difference which leads to a significant enhancement of CP violating
observables.

The decay amplitudes of a $\tau$ lepton into three pions,
$\tau\rightarrow (3\pi)\nu_\tau$, are dominated by contributions of
the two overlapping resonances with different spins and
relatively large width-to-mass ratios\cite{Tauola} :
\begin{eqnarray}
\begin{array}{clll}
 a_1 : & {\rm J}^{\rm P}=1^+, & m_{a_1}=1.251\ \ {\rm  GeV},
       & \ \Gamma_{a_1}=0.599\ \ {\rm  GeV},\\
 \pi^\prime : & {\rm J}^{\rm P}=0^-, & m_{\pi^\prime}=1.300\ \ {\rm GeV},
       & \ \Gamma_{\pi^\prime}= 0.300\ \ {\rm  GeV}.
\label{TwoR}
\end{array}
\end{eqnarray}
Here we should note that even the parameters of the $a_1$ are not
accurately determined.  In the $3\pi$ decay mode of the $\tau$ various
phenomenological parametrizations\cite{Bowler,Isgur,Decker,Tauola}
of the form factors have been used to analyze experimental
data\cite{Ruckstuhl,Albrecht}.
We adopt the parametrization of the $\tau$-decay
library, TAUOLA\cite{Tauola}, where the $\pi^\prime$ parameters given in
eq.~(\ref{TwoR}) can also be found.

The three-charged-pion decay mode of the $\tau$ is promising
for the detection of  CP violation due to the following reasons.
First, no tagging of the other $\tau$ is necessary.
Second, the three-charged-pion mode can be measured not only at the
conventional machines (e.g. CESR and LEP\cite{Gomez}) but also
at the planned $B$ factories and at $\tau$-Charm
factories\cite{Pich} where many $\tau$
leptons (yearly $10^7$ to $10^8$) are expected to be produced.
Third, reconstruction of the $\tau$ rest frame is easy for
the three-prong decay modes. Since at least two neutrinos escape
detection it is in general difficult to reconstruct the $\tau$ rest
frame in $\tau^+\tau^-$ production. There are a few situations
where the rest frame of the $\tau$ can actually be reconstructed.
One is $\tau$-pair production close to threshold where $\tau$ leptons
are produced at rest. This possibility can be realized at future
$\tau$-Charm factories\cite{Pich}. Another is when both $\tau$ leptons
decay into hadrons. In this case impact-parameter methods\cite{Kuhn}
allow us to reconstruct the rest frame of the $\tau$ even for $\tau$'s
in flight.  However, in the three-charged-pion decay mode,
the direction of the $\tau$ can be directly reconstructed through
the precise and simultaneous determination of the $\tau$ production
and decay points.

On the other hand, it is often difficult to make a reliable
quantitative prediction for CP violation in hadronic
$\tau$- decay modes due to the uncertainty in the hadronic matrix
elements. This problem can, however, be removed by considering final
states dominated by at least two neighboring resonances.
The precise experimental determination of the widths and masses of
the resonances will give a reliable handle on the calculation of CP
asymmetries.
In fact, the parametrization of the three-pion decay
mode has been investigated quite thoroughly by many authors.
Exhaustive discussion of the decay $\tau\rightarrow 3\pi\nu_\tau$
is given in Refs.\ \cite{Santa,Mirkes,Isgur}.
A complete parametrization of
the $a_1/\pi^\prime\rightarrow 3\pi$ decay currents can be found in
the TAUOLA, which adopts from Ref.\ \cite{Isgur}
$f_{\pi^\prime}=0.02\sim 0.08$GeV
However, the quoted value of $f_{\pi^\prime}$
may be overestimated because the mixing between the chiral pion field
and a massive pseudoscalar $q\bar{q}$ bound state should
be considered. Since any CP-violating effect in the decay
$\tau\rightarrow 3\pi\nu_\tau$ depends on $f_{\pi^\prime}$,
it is crucial to estimate its magnitude. We devote one section to this issue.

The paper is organized as follows.
In section~2 the decay distribution for $\tau\rightarrow (3\pi)\nu_\tau$
is presented. Its general form and our resonance parametrizations are
given explicitly. In section 3 CP-violating asymmetries in the decay
$\tau\rightarrow 3\pi\nu_\tau$  are introduced.
In section~4 the multi-Higgs-doublet (MHD) model\cite{Branco,Grossman}
and scalar-leptoquark (SLQ) models\cite{Hall} are introduced as examples of
models which can generate CP-violating effects
in the decay $\tau\rightarrow 3\pi\nu_\tau$.
In section 5 we estimate the magnitude of $f_{\pi^\prime}$ by
employing a simplified chiral Lagrangian coupled to a
massive pseudoscalar field. Finally, in section 6 the possibility
to detect CP violation in the three charged pion decay mode of
the $\tau$ is discussed quantitatively. Section 7 summarizes our findings.

\section{Distributions}
\cleqn

\pr
The matrix element for the decay $\tau^-\rightarrow (3\pi)^-\nu_\tau$
is written in the form:
\begin{eqnarray}
M=\sqrt{2}G_F \Bigl[(1+\chi)\bar{u}(k,-)P_+\gamma^\mu u(p,\sigma)J_\mu
                 +\eta\bar{u}(k,-)P_+ u(p,\sigma)J_P\Bigr],
\label{decaym}
\end{eqnarray}
with  $P_\pm=(1\pm \gamma_5)/2$.
Here $G_F$ is the Fermi constant; $p$ and $k$ are the four-momenta
of the $\tau$ lepton and the tau neutrino, respectively. $\chi$ and
$\eta$ are complex numbers parametrizing the contribution from physics
beyond the SM.
The spin-quantization direction of the $\tau^-$ is taken to be
the direction opposite to the neutrino momentum (See Fig.~1(a).)
and its helicity is denoted by $\sigma$ ($\sigma=\pm 1$).
The tau neutrino is left-handed, so its helicity is $-1$ as
indicated by $u(k,-)$.
$J_\mu$ and $J_P$ are the vector and scalar hadronic matrix elements,
respectively, and are given by
\begin{eqnarray}
J_\mu= <(3\pi)^-|\bar{d}\gamma_\mu (1-\gamma_5) u|0>,\qquad
J_P= <(3\pi)^-|\bar{d}(1+\gamma_5)u|0>.
\end{eqnarray}

Both the $a_1$ and $\pi^\prime$ resonances contribute to $J_\mu$
while only the pseudoscalar state, $\pi^\prime$, contributes
to $J_P$. The explicit form of $J_\mu$  is found e.g. in the TAUOLA,
and the hadronic (pseudo)-scalar current, $J_P$, is determined from
the Dirac equation via the expression for $J_\mu$:
\begin{eqnarray}
&&J_\mu=N\Bigl\{\frac{2\sqrt{2}}{3}T^{\mu\nu}
   \bigl[(q_2-q_3)_\nu F_1(q^2,s_1)
   +(q_1-q_3)_\nu F_2(q^2,s_2)\bigr]\nonumber\\
   &&\hskip 1.3cm +q^\mu C_{\pi^\prime}\bigl[s_1(s_2-s_3)F_3(q^2,s_1)
      +s_2(s_1-s_3)F_4(q^2,s_2)\bigr]\Bigr\},
 \label{Hcurrent}  \\
&&J_{P\pm}=\frac{1}{m_u+m_d}
        q^\mu J_\mu\nonumber\\
&&\hskip 0.7cm \approx\frac{m^2_{\pi^\prime}}{m_u+m_d}
        NC_{\pi^\prime}\bigl[s_1(s_2-s_3)F_3(q^2,s_1)
        +s_2(s_1-s_3)F_4(q^2,s_2)\bigr].
\end{eqnarray}
Here $q_1$ and $q_2$ are the four-momenta of two identical $\pi^-$'s,
$q_3$ is the four-momentum of the $\pi^+$, and $q$ is the four-momentum
of the $(3\pi)^-$ system in the decay of the $\tau^-$.
$m_u$ and $m_d$ are the current
masses of the $u$ and $d$ quarks, respectively.
The invariant mass squared of the $3\pi$ system, $q^2$, and the
three kinematic invariants, $s_i$ ($i=1,2,3$), are defined in terms
of the three pion momenta, $q_i$ ($i=1,2,3$), as
\begin{eqnarray}
q^2=(q_1+q_2+q_3)^2,\qquad s_1=(q_2+q_3)^2,\qquad
s_2=(q_3+q_1)^2,\qquad s_3=(q_1+q_2)^2.
\end{eqnarray}
Also, $T^{\mu\nu}=g^{\mu\nu}-q^\mu q^\nu/q^2$.
The form factors, $F_i$ ($i=1$ to $4$), are normalized
such that $F_i(0,0)=1$, and the coefficient $2\sqrt{2}/3$ in (\ref{Hcurrent})
is fixed by the soft pion theorem.
TAUOLA uses the parameters
\begin{eqnarray}
N=\frac{\cos\theta_C}{f_\pi}, \qquad
C_{\pi^\prime}=\frac{g_{\pi^\prime\rho\pi}g_{\rho\pi\pi}
                       f_{\pi^\prime}f_\pi}{m^4_\rho m^2_{\pi^\prime}},
\end{eqnarray}
and adopts the parameter values
\begin{eqnarray}
&&\cos\theta_C=0.973, \qquad m_\rho=0.773\ \ {\rm GeV},\nonumber\\
&&f_\pi=0.0933\ \ {\rm GeV},\qquad f_{\pi^\prime}=0.02\ \
   {\rm GeV},\nonumber\\
&& g_{\pi^\prime\rho\pi}=5.8, \qquad g_{\rho\pi\pi}=6.08.
\label{ParaV}
\end{eqnarray}

It is convenient to cast the decay amplitude (\ref{decaym}) into the form
\begin{eqnarray}
M=\sqrt{2}G_F(1+\chi)\left[\sum_\lambda L_{\sigma\lambda}H_\lambda
  +(1+\xi)L_{\sigma s}H_s\right],
\label{newm}
\end{eqnarray}
where the parameter $\xi$ is given in terms of the $\chi$
and $\eta$ in eq.~(\ref{decaym}) by
\begin{eqnarray}
\xi=\frac{m^2_{\pi^\prime}}{(m_u+m_d)m_\tau}
    \left(\frac{\eta-\chi}{1+\chi}\right).
\end{eqnarray}
The $\tau^-\rightarrow \nu_\tau$ transition amplitudes,
$L_{\sigma\lambda}$ ($\lambda=0,\pm$) and $L_{\sigma s}$, in
eq.~(\ref{newm}) are defined as
\begin{eqnarray}
&&L_{\sigma\lambda}=\bar{u}(k,-)\gamma^\mu P_- u(p,\sigma)
   \epsilon^*_\mu(q,\lambda),\nonumber\\
&&L_{\sigma s}=\bar{u}(k,-)P_+ u(p,\sigma),
\end{eqnarray}
where the polarization vector $\epsilon(q,\lambda)$ of the virtual
vector meson $a_1^-$ satisfies
\begin{eqnarray}
\sum_{\lambda=0,\pm}\epsilon^\mu(q,\lambda)\epsilon^{*\nu}(q,\lambda)
 = -g^{\mu\nu}+\frac{q^\mu q^\nu}{q^2}.
\end{eqnarray}
The purely leptonic amplitudes are known functions of the kinematic
variables and are given in the convention of Ref.~\cite{Hagiwara1} by
\begin{eqnarray}
L_{\sigma +}&=&0,\nonumber\\
L_{\sigma 0}&=&\frac{m_\tau}{\sqrt{q^2}}\sqrt{m^2_\tau-q^2}
                 \delta_{\sigma +},\nonumber\\
L_{\sigma -}&=&\sqrt{2}\sqrt{m^2_\tau-q^2}
                 \delta_{\sigma -},\nonumber\\
L_{\sigma s}&=&\sqrt{m^2_\tau-q^2}\delta_{\sigma +}.
\end{eqnarray}
The hadronic matrix elements, $H_\lambda$ ($\lambda=0,\pm$) and $H_s$,
are
\begin{eqnarray}
&&H_\lambda=-\frac{2\sqrt{2}}{3}N\epsilon_\mu(q,\lambda)
            [(q_2-q_3)^\mu F_1(q^2,s_1)+(q_1-q_3)^\mu F_2(q^2,s_2)],\\
&&H_s=Nm_\tau C_{\pi^\prime}
              [s_1(s_2-s_3)F_3(q^2,s_1)+s_2(s_1-s_3)F_4(q^2,s_2)].
\label{Had1}
\end{eqnarray}

Assuming single meson dominance in each channel TAUOLA expresses
the form factors $F_i$ ($i=1$ to $4)$ in terms of the
meson propagators as
\begin{eqnarray}
F_1(q^2,s_1)&=&B_{a_1}(q^2)B_\rho(s_1), \qquad
F_2(q^2,s_2)=B_{a_1}(q^2)B_\rho(s_2), \nonumber\\
F_3(q^2,s_1)&=&B_{\pi^\prime}(q^2)B_\rho(s_1), \qquad
F_4(q^2,s_2)=B_{\pi^\prime}(q^2)B_\rho(s_2).
\end{eqnarray}
The $a_1$, $\pi^\prime$, and $\rho$ meson propagators are
parametrized\cite{Tauola} in the Breit-Wigner
form with momentum-dependent widths:
\begin{eqnarray}
B_{X}(q^2)=\frac{m^2_X}{m^2_X-q^2-im_X\Gamma_X(q^2)},\qquad
\Gamma_X(q^2)=\Gamma_X\left(\frac{f_X(q^2)}{f_X(m^2_X)}\right),
\end{eqnarray}
for $X=a_1$, $\pi^\prime$ or $\rho$.
The momentum-dependence of all widths has been determined
from experimental data. The parametrizations of the widths
available from the TAUOLA are:
\begin{eqnarray}
&&f_{a_1}(q^2)=\left\{\begin{array}{l}
  q^2[1.623+10.38/q^2-9.32/q^4+0.65/q^6],\ \
  {\rm for}\ \ q^2>(m_\rho+m_\pi)^2\\
                               { }       \\
  4.1(q^2-9m^2_\pi)^3[1-3.3(q^2-9m^2_\pi)+5.8(q^2-9m^2_\pi)^2],
  \ \ {\rm elsewhere},
               \end{array}\right. \\
&& { }\nonumber\\
&&f_{\pi^\prime}(q^2)=\left\{\begin{array}{cl}
    \frac{m^2_{\pi^\prime}}{q^2}
    \left(\frac{P_\pi(q^2)}{P_\pi(m^2_\rho)}\right)^5,&
    \ \ {\rm for}\ \ q^2>(m_\rho+m_\pi)^2\\
        { } & { } \\
    0, & \ \ {\rm elsewhere},
              \end{array}\right.
  \label{pi1_f}\\
&& { }\nonumber\\
&&f_{\rho}(q^2)=\left\{\begin{array}{cl}
     \frac{m_\rho}{\sqrt{q^2}}
       \left(\frac{q^2-4m^2_\pi}{m^2_\rho-4m^2_\pi}\right)^{3/2},
        & \ \ {\rm for}\ \ q^2>(2m_\pi)^2\\
          { } & { }\\
      0, &\ \ {\rm elsewhere}.
               \end{array}\right.
\end{eqnarray}
$P_\pi$ in eq.~(\ref{pi1_f}) denotes the momentum of the $\pi$
in the (virtual) $\pi^\prime$
rest frame.
In addition to the parameters  (\ref{TwoR}) and  (\ref{ParaV})
TAUOLA adopts for the $\rho$ meson width
\begin{eqnarray}
\Gamma_{\rho}=0.145\ \ {\rm GeV}.
\end{eqnarray}

The amplitude for the $\tau^+$ decay into $(3\pi)^+$, which is the
CP-conjugated process of the $\tau^-$ decay into $(3\pi)^-$,
can be determined in the same manner as that of the $\tau^-$ decay
into $(3\pi)^-$. For the sake of discussion we use the same kinematic
variables as in the $\tau^-$ decay. In the $\tau^+$ decay
$q_1$ and $q_2$ are the four-momenta of two identical $\pi^+$'s,
$q_3$ is the four-momentum of the $\pi^-$, and $q$ is the four-momentum
of the $(3\pi)^+$ system. The decay amplitude $\bar{M}$ for the process
$\tau^+\rightarrow(3\pi)^+\bar{\nu}_\tau$ can be written in the form
\begin{eqnarray}
\bar{M}=\sqrt{2}G_F(1+\bar{\chi})\left[\sum_\lambda \bar{L}_{\sigma\lambda}
  \bar{H}_\lambda +(1+\bar{\xi})\bar{L}_{\sigma s}\bar{H}_s\right].
\label{newp}
\end{eqnarray}
The $\tau^+\rightarrow \bar{\nu}_\tau$ transition amplitudes,
$\bar{L}_{\sigma\lambda}$ ($\lambda=0,\pm$) and $\bar{L}_{\sigma s}$,
are given by
\begin{eqnarray}
&&\bar{L}_{\sigma\lambda}=\bar{v}(p,\sigma)\gamma^\mu P_- v(k,+)
   \epsilon^*_\mu(q,\lambda),\nonumber\\
&&\bar{L}_{\sigma s}=\bar{v}(p,\sigma)P_+ v(k,+),
\end{eqnarray}
where $\epsilon(q,\lambda)$ is the polarization vector of the virtual
vector meson $a_1^+$ with the helicity value $\lambda$.
We note that the hadronic amplitudes, $\bar{H}_{\lambda}$ ($\lambda=0,\pm$)
and $\bar{H}_s$, for the decay $\tau^+\rightarrow (3\pi)^+\bar{\nu}_\tau$
are the same as the hadronic amplitudes $H_\lambda(\lambda=0,\pm)$
and $H_s$ for the decay $\tau^-\rightarrow (3\pi)^-\nu_\tau$:
\begin{eqnarray}
\bar{H}_\lambda = H_\lambda,\qquad  \bar{H}_s = H_s.
\label{Had2}
\end{eqnarray}
Hence it is straightforward to obtain each amplitude of the $\tau^+$ decay
into $(3\pi)^+$ from the corresponding $\tau^-$ decay amplitude
(\ref{decaym}) through the CP relations among the amplitudes and
the couplings
\begin{eqnarray}
&&\bar{L}_{\sigma\lambda}=(-1)^{\lambda}L_{-\sigma,-\lambda},\qquad
  \bar{L}_{\sigma s}=L_{-\sigma s},\\
&&\bar{\chi}=\chi^*,\qquad \bar{\eta}=\eta^*,
\end{eqnarray}
in the approximation where the imaginary parts of the
intermediate $W$ and scalar propagators are neglected.

The decay of the $\tau^\pm$ into three charged pions depends on
five independent kinematic variables. Because of two intermediate
resonances, $a_1^\pm$ and $\pi^{\prime\pm}$,
decaying into $(3\pi)^\pm$ it is convenient to consider
both the $\tau^\pm$ rest frame and the $(3\pi)^\pm$ rest
frame. We define two coordinate systems as shown in Fig.~1.
The two coordinate systems have a common $y$-axis
which is chosen along the $\vec{k}\times \vec{q}_3$ direction.
Here $\vec{k}$ is the three-momentum of the tau neutrino.
In the ($x,y,z$) coordinate system the $z$-axis is along the direction
of the tau neutrino momentum, $\vec{k}$, and the momentum $\vec{q}_3$
is in the postive-$x$ half-plane. The starred coordinate system is defined
with a rotation by $\theta$, the angle between $\vec{k}$ and $\vec{q_3}$,
with respect to the common $y$-axis from the $(x,y,z)$ coordinate system.
We now consider the five variables; $q^2$, $s_1$, $s_2$,
$\theta$, and the azimuthal angle $\phi^*$ of the
$\pi^\pm$ momentum $\vec{q}_1$
with respect to the $x^*$-axis in the starred coordinate system
(See Fig.~1(b).)

The range of the variables is given by
\begin{eqnarray}
&&(3m_\pi)^2\leq q^2 \leq m^2_\tau,\qquad
  (2m_\pi)^2\leq s_1 \leq (\sqrt{q^2}-m_\pi)^2,\nonumber\\
&&s_{2min} \leq s_2\leq s_{2max},\qquad 0\leq \theta \leq \pi,\qquad
  0\leq \phi^* \leq 2\pi,
\end{eqnarray}
where for given $q^2$ and $s_1$ the minimum and maximum values of $s_2$
values are
\begin{eqnarray}
&&s_{2min}=\frac{1}{2}\left[q^2+3m^2_\pi-s_1
   -\sqrt{(1-4m^2_\pi/s_1)}
    \cdot\lambda^{1/2}(q^2,m^2_\pi,s_1)\right],\nonumber\\
&&s_{2max}=\frac{1}{2}\left[q^2+3m^2_\pi-s_1
   +\sqrt{(1-4m^2_\pi/s_1)}\cdot\lambda^{1/2}(q^2,m^2_\pi,s_1)\right],
\end{eqnarray}
with the well-known triangle function,
$\lambda(x,y,z)=x^2+y^2+z^2-2(xy+yz+zx)$.
For simplicity we introduce the following notations:
\begin{eqnarray}
&&\lambda_i=\lambda(q^2,m^2_\pi,s_i)\ \ (i=1,2,3),\qquad
  G=\sqrt{-\lambda(\lambda_1,\lambda_2,\lambda_3)}.
\end{eqnarray}
Using the kinematic variables and notations we obtain analytic
expressions of the hadronic decay amplitudes (\ref{Had1}):
\begin{eqnarray}
&&H_\pm=\mp\frac{N}{6\sqrt{q^2\lambda_3}}B_{a_1}(q^2)
    \left[A\sin\theta+B(\cos\theta\cos\phi^*\mp i\sin\phi^*)\right],
    \label{Hpm}\\
&&H_0=-\frac{\sqrt{2}N}{6\sqrt{q^2\lambda_3}}
        B_{a_1}(q^2)\left[A\cos\theta -B\sin\theta\cos\phi^*\right],
    \label{H0}\\
&&H_s=\frac{Nm_\tau C_{\pi^\prime}}{2}
        B_{\pi^\prime}(q^2)\left[C\right],
    \label{Hs}
\end{eqnarray}
where
\begin{eqnarray}
&&A= 3\lambda_3[B_\rho(s_1)+B_\rho(s_2)]
    -(\lambda_1-\lambda_2)[B_\rho(s_1)-B_\rho(s_2)],\nonumber\\
&&B=G[B_\rho(s_1)-B_\rho(s_2)],\nonumber\\
&&C=2s_1(s_2-s_3)B_\rho(s_1)+2s_2(s_1-s_3)B_\rho(s_2).
\end{eqnarray}
The hadronic amplitudes, $H_\lambda$ ($\lambda=0,\pm$) and $H_s$, are
symmetric under the interchange of two identical pions.
The functions $A$ and $C$
also are symmetric, while the function $B$ is antisymmetric.

We denote the differential decay rates of the $\tau^\mp$ into three pions
as
\begin{eqnarray}
&&G(q^2,s_1,s_2,\cos\theta,\phi^*)
=\frac{{\rm d}^5}{{\rm d}q^2
 {\rm d}s_1{\rm d}s_2{\rm d}\cos\theta{\rm d}\phi^*}
\Gamma[\tau^-\rightarrow (3\pi)^-\nu_\tau],\nonumber\\
&&\bar{G}(q^2,s_1,s_2,\cos\theta,\phi^*)
= \frac{{\rm d}^5}{{\rm d}q^2
 {\rm d}s_1{\rm d}s_2{\rm d}\cos\theta{\rm d}\phi^*}
\bar{\Gamma}[\tau^+\rightarrow(3\pi)^+\bar{\nu}_\tau],
\end{eqnarray}
and examine the consequences of CP invariance in the $\tau^\mp$
differential decay rates.
The CP transformation should relate the decay
$\tau^-\rightarrow (3\pi)^-\nu_\tau$
to the decay $\tau^+\rightarrow (3\pi)^+\bar{\nu}_\tau$.
We find that CP invariance leads to the following relation
in the differential decay rates:
\begin{eqnarray}
&&G(q^2,s_1,s_2,\cos\theta,\phi^*)
  \stackrel{\rm CP}{=}
  \bar{G}(q^2,s_1,s_2,\cos\theta,-\phi^*).
\end{eqnarray}
The relation enables us to construct a CP-conserving sum, $\Sigma$,
and a CP-violating difference, $\Delta$, of the differential $\tau^\pm$
decay rates:
\begin{eqnarray}
&&\Sigma=G(q^2,s_1,s_2,\cos\theta,\phi^*)
    +\bar{G}(q^2,s_1,s_2,\cos\theta,-\phi^*),\nonumber\\
&&\Delta=G(q^2,s_1,s_2,\cos\theta,\phi^*)
    -\bar{G}(q^2,s_1,s_2,\cos\theta,-\phi^*).
\end{eqnarray}
It is rather straightforward to obtain from the amplitudes (\ref{newm})
and (\ref{newp}) the analytic forms of
$\Sigma$ and $\Delta$;
\begin{eqnarray}
&&\Sigma=F(q^2)\!\left\{2|H_-|^2\!+\!\frac{m^2_\tau}{q^2}|H_0|^2
    +|1+\xi|^2|H_s|^2\!-\!2\frac{m_\tau}{\sqrt{q^2}}
         \Bigl[1\!+\!{\rm Re}(\xi)\Bigr]
         \Bigl[{\rm Re}(H_0H^*_s)\Bigr]\right\},
   \label{Sigma}\nonumber\\
&&\Delta=-2F(q^2)\frac{m_\tau}{\sqrt{q^2}}
     \Bigl[{\rm Im}(\xi)\Bigr]\Bigl[{\rm Im}(H_0H^*_s)\Bigr],
\label{Acp}
\end{eqnarray}
where for notational convenience we have introduced the overall
function $F(q^2)$
\begin{eqnarray}
F(q^2)=\frac{G_F^2m_\tau}{2^{7}\pi^6}\frac{(1-q^2/m^2_\tau)^2}{q^2}
       |1+\chi|^2.
\end{eqnarray}
{}From the expression (\ref{Acp}) we note two important features of
the CP-violating distribution $\Delta$;
(i) Every CP-violating asymmetry requires not only a non-vanishing
    ${\rm Im}(\xi)$ but also requires the interference between the
    longitudinal $a_1$ mode and the $\pi^\prime$ mode, and
(ii) the difference $\Delta$ depends only on $\cos\phi^*$, but it does
    not depend on $\sin\phi^*$ which is odd under the naive time reversal
    $\tilde{\rm T}$ (See eqs.~(\ref{H0}) and (\ref{Hs})).
Here the transformation $\tilde{\rm T}$ means $t\rightarrow -t$
without the interchange of initial and final states.

\section{Asymmetries}
\cleqn

\pr
As shown in section~2 every CP asymmetry
requires not only a non-vanishing ${\rm Im}(\xi)$ but also the interference
between the longitudinal $a_1$ mode and the pseudoscalar $\pi^\prime$ mode;
the interference is proportional to the $f_{\pi^\prime}$.
More explicitly $\Delta$ is proportional to the product of $f_{\pi^\prime}$
and ${\rm Im}(\xi)$,
\begin{eqnarray}
\Delta\propto\left[{\rm Im}(\xi)\right]\cdot
       \left[f_{\pi^\prime}\right].
\end{eqnarray}

In order to observe $\Delta$ it is useful
to form an observable with an appropriate real weight function
$w(q^2,s_1,s_2,\cos\theta,\phi^*)$. A CP-violating scalar quantity is
then obtained as
\begin{eqnarray}
\langle w\Delta\rangle=\int\left[w(q^2,s_1,s_2,\cos\theta,\phi^*)
    \cdot\Delta\right]
    {\rm d}q^2{\rm d}s_1{\rm d}s_2{\rm d}\cos\theta{\rm d}\phi^*,
\end{eqnarray}
where $\langle X\rangle$ means the integration of the quantity $X$ over
the allowed phase space
of $q^2$, $s_1$, $s_2$, $\cos\theta$, and $\phi^*$.
The statistical significance of this observable can be determined by
the quantity
\begin{eqnarray}
\varepsilon
   =\frac{\langle w\Delta\rangle}{\sqrt{\langle\Sigma\rangle
     \cdot\langle w^2\Sigma\rangle}},
\label{Significance}
\end{eqnarray}
with $\Sigma$ given in eq.~(\ref{Sigma}).
The number of $\tau$ leptons required
to observe the effect at the $1$-$\sigma$ level is then
\begin{eqnarray}
N_\tau=\frac{1}{{\rm Br}\cdot\varepsilon^2},
\end{eqnarray}
where Br denotes the branching ratio of the $\tau$ decay into three
charged pions, which is 6.8\%\cite{Albrecht}.
By appropriately choosing the weight function, $w$, the CP asymmetry
$\langle w\Delta\rangle$ can be made large.
The optimal weight function maximizing the quantity $\varepsilon$
in eq.~(\ref{Significance}) is known\cite{Soni3} to be
\begin{eqnarray}
w_{opt}(q^2,s_1,s_2,\cos\theta,\phi^*)
 =
\frac{\Delta(q^2,s_1,s_2,\cos\theta,\phi^*)}{\Sigma(q^2,s_1,s_2,
      \cos\theta,\phi^*)}.
\end{eqnarray}

We can also consider an observable with $w=\pm 1$ in phase space, which
corresponds to the usual definition of an asymmetry.
The distribution property of $\Delta$ in eq.~(\ref{Acp})
suggests that we should consider two types of CP-violating
forward-backward asymmetries, $A_{1FB}$ and $A_{2FB}$,
whose weight functions are given by
\begin{eqnarray}
&&w_1(q^2,s_1,s_2,\cos\theta,\phi^*)={\rm sign}[\cos\theta],\nonumber\\
&&w_2(q^2,s_1,s_2,\cos\theta,\phi^*)
       =
  {\rm sign}[s_2-s_1]\cdot{\rm sign}[\cos\phi^*],
\end{eqnarray}
respectively.
Note that the factor ${\rm sign}[s_2-s_1]$ is included in the definition of
the $A_{2FB}$ asymmetry because the CP-violating asymmetry with only
${\rm sign}[\cos\phi^*]$ as a weight function vanishes due to Bose
symmetry.

\section{Models}
\cleqn

\pr
In the SM CP violation arises from a nontrivial phase
in the KM flavor-mixing matrix\cite{Kobayashi}
in the hadronic charged current, but the KM-type CP violation
can not be detected in $\tau$ decays.
As possible new sources of CP violation detectable in the $\tau$ decay
we consider new scalar-fermion interactions which preserve
the symmetries of the SM.
Then, it can be proven that only four types of
scalar-exchange models\cite{Davies} contribute to the decay
$\tau\rightarrow (3\pi)\nu_\tau$.
One of them is the multi-Higgs-doublet (MHD) model\cite{Grossman,Branco}
and the other three models are scalar-leptoquark (SLQ)
models\cite{Buch,Hall}.

\subsection{Multi-Higgs-doublet (MHD) models}

\pr
In this subsection we consider a MHD model with $n$ Higgs doublets.
The Yukawa interaction of the MHD model is
\begin{eqnarray}
{\cal L}_{MHD}&=&\bar{Q}_{L_i}F^D_{ij}\Phi_d D_{R_j}+
     +\bar{Q}_{L_i}F^U_{ij}\tilde{\Phi}_u U_{R_j}
     +\bar{L}_{L_i}F^E_{ij}\Phi_e E_{R_j}+{\rm h.c.}
\end{eqnarray}
Here $Q_{L_i}$ denotes left-handed quark doublets, and $L_{L_i}$
denotes left-handed lepton doublets.
$D_{R_i}$ ($U_{R_i}$) and $E_{R_i}$ are for right-handed
down (up) quark singlets and  right-handed charged lepton
singlets, respectively.
The sub-index $i$ is a generation index ($i=1,2,3$).
$\Phi_j$ $(j=1$ to $n)$ are $n$ Higgs doublets and
$\tilde{\Phi}_j=i\sigma_2\Phi^*_j$.
Sub-indices $d$, $u$ and $e$ denote the Higgs doublets that couple to
down-type quarks, up-type quarks and charged leptons, respectively.
$F^U$ and $F^D$ are general $3\times 3$ Yukawa matrices of which one
matrix can be taken to be real and diagonal. Since neutrinos are
massless $F^E$ can be chosen real and diagonal.
The MHD model has $2(n-1)$ charged and $2n-1$ neutral
physical scalars, and the Yukawa interactions of the $2(n-1)$ physical
charged scalars with fermion mass eigenstates read
\begin{eqnarray}
{\cal L}_{MHD}=\sqrt{2\sqrt{2}G_F}\sum_{i=2}^n
         \left[X_i(\bar{U}VM_DD_R)+Y_i(\bar{U}_RM_UVD_L)
        +Z_i(\bar{N}_LM_EE_R)\right]H^+_i+{\rm h.c.}\nonumber\\
       { }
\end{eqnarray}
Here $M_D$, $M_U$ and $M_E$ denote the diagonal mass matrices of down-type
quarks, up-type quarks and charged leptons, respectively.
$H^+_i$  the positively charged Higgs particles. $N_L$ are for
left-handed neutrino fields and $V$ for the KM matrix.
$X_i$, $Y_i$ and $Z_i$ are complex coupling constants which arise from
the mixing matrix for charged scalars.

Within the framework of the MHD model, CP violation in charged scalar
exchange can arise for more than two Higgs doublets\cite{Weinberg,Glashow}.
There are two mechanisms which give rise to CP violation in the scalar
sector. In one mechanism\cite{Lee,Weinberg} CP symmetry is maintained at
the Lagrangian level but broken through complex vacuum expectation values.
However, this possibility has been shown to have some phenomenological
difficulties\cite{Pokorski,Grossman}.
In the other mechanism CP is broken by complex Yukawa couplings
and possibly by complex vacuum expectation values so that CP violation can
arise from both charged scalar exchange and from $W^\pm$ exchange.
CP violation in both mechanisms
is commonly manifest in phases that appear in the combinations
$XY^*$, $XZ^*$ and $YZ^*$.

One crucial condition for CP violation in the MHD model is
that not all the charged scalars should be degenerate.
Then, without loss of generality and for simplicity, we can assume that
all but the lightest of the charged scalars effectively decouple from
fermions. The couplings of the lightest charged scalar to fermions
are described by a simple Lagrangian
\begin{eqnarray}
{\cal L}_{MHD}=\sqrt{2\sqrt{2}G_F}
         \left[X(\bar{U}VM_DD_R)+Y(\bar{U}_RM_UVD_L)
        +Z(\bar{N}_LM_EE_R)\right]H^++{\rm h.c.}
\label{MHD_L}
\end{eqnarray}
This Lagrangian gives the effective Lagrangian for the decay
$\tau\rightarrow 3\pi\nu_\tau$
\begin{eqnarray}
{\cal L}^{MHD}_{eff}=2\sqrt{2}G_FV^*_{ud}m_\tau
          \left[m_d\frac{X^*Z}{M^2_{H}}
          (\bar{u}_Ld_R)(\bar{\nu}_{\tau_L}\tau_R)
         +m_u\frac{Y^*Z}{M^2_{H}}
            (\bar{u}_Rd_L)(\bar{\nu}_{\tau_L}\tau_R)\right],
\end{eqnarray}
at energies which are low compared to the mass of the charged Higgs boson.
Then one can show that the contribution from the MHD model in the
$\tau\rightarrow 3\pi\nu_\tau$ decay distribution of eq.~(\ref{decaym})
is represented by the parameters
\begin{eqnarray}
\chi_{MHD}=0,\qquad \eta_{MHD}=\frac{m_d}{M^2_{H}}
            \left[X^*Z-\left(\frac{m_u}{m_d}\right)Y^*Z\right],
\end{eqnarray}
and  CP violation in the MHD model is determined by the parameter
\begin{eqnarray}
{\rm Im}(\xi_{MHD})=-\left(\frac{m_d}{m_u+m_d}\right)
                     \left(\frac{m^2_{\pi^\prime}}{M^2_H}\right)
    \left[{\rm Im}(XZ^*)-\left(\frac{m_u}{m_d}\right){\rm Im}(YZ^*)\right].
\label{MHDp}
\end{eqnarray}

\subsection{Scalar-leptoquark (SLQ) models}

\pr
In this subsection we discuss CP-violating effects from leptoquark exchange.
There are three types of SLQ models\cite{Davies,Buch}
which can contribute to the decay $\tau\rightarrow 3\pi\nu_\tau$ at the
tree level. The quantum numbers of the three leptoquarks under the gauge
group SU(3)$_{\rm C}\times$SU(2)$_{\rm L}\times$U(1)$_{\rm Y}$ are
\begin{eqnarray}
&&\Phi_1=\left(3,2,\ \ \frac{7}{6}\right) \ \
         ({\rm model}\ \ {\rm I}),\nonumber\\
&&\Phi_2=\left(3,1,-\frac{1}{3}\right) \ \ ({\rm model}\ \ {\rm II}),
  \nonumber\\
&&\Phi_3=\left(3,3,-\frac{1}{3}\right) \ \ ({\rm model}\ \ {\rm III}),
\end{eqnarray}
respectively. The hypercharge $Y$ is defined to be
$Q=I_3+Y$. The Yukawa couplings of the leptoquarks to fermions are
given by
\begin{eqnarray}
&&{\cal L}^I_{SLQ}=[-x_{ij}\bar{Q}_{L_i}i\tau_2 E_{R_j}
      +x^\prime_{ij}\bar{U}_{R_i}L_{L_j}]\Phi_1+{\rm h.c.},\nonumber\\
&&{\cal L}^{II}_{SLQ}=[y_{ij}\bar{Q}_{L_i}i\tau_2L^c_{L_j}
      +y^\prime_{ij}\bar{U}_{R_i}E^c_{R_j}]\Phi_2+{\rm h.c.},\nonumber\\
&&{\cal L}^{III}_{SLQ}=z_{ij}[\bar{Q}_{L_i}i\tau_2\vec{\tau}L^c_{L_j}]
       \cdot\vec{\Phi}_3+{\rm h.c.}
\end{eqnarray}
Here the coupling constants $x^{(\prime)}_{ij}$, $y^{(\prime)}_{ij}$ and
$z_{ij}$ are complex when CP violation arises from the Yukawa interactions.
$\bar{Q}_{L_i}=(\bar{u}_i,\bar{d}_i)_L$ and
$L_{L_i}=(\bar{\nu}_i,\bar{e}_i)_L$.
The superscript $c$ denotes charge conjugation, i.e.
$\psi^c_{R,L}=i\gamma^0\gamma^2\bar{\psi}^T_{R,L}$ for a spinor field
$\psi$. $\vec{\tau}=(\tau_1,\tau_2,\tau_3)$ and $\tau_i$ ($i=1,2,3$)
are the Pauli matrices.
In terms of the charge component of the leptoquarks, the
Lagrangian relevant to the $\tau\rightarrow 3\pi\nu_\tau$ decay is given by
\begin{eqnarray}
&&{\cal L}^I_{SLQ}=\left[x_{13}\bar{d}_{L}\tau_{R}+x^\prime_{13}
                  \bar{u}_{R}\nu_{\tau_L}\right]\phi^{(2/3)}_1
                  +{\rm h.c.},\nonumber\\
&&{\cal L}^{II}_{SLQ}=\left[-y_{13}(\bar{u}_{L}\tau^c_{L}
                   -\bar{d}_{L}\nu^c_{\tau_L})
                  + y^\prime_{13}\bar{u}_R\tau^c_{R}\right]\phi^{(-1/3)}_2
                  + {\rm h.c.},\nonumber\\
&&{\cal L}^{III}_{SLQ}=-z_{13}[\bar{u}_{L}\tau^c_{L}
                  +\bar{d}_L\nu^c_{\tau_L}]
                  \phi^{(-1/3)}_3+{\rm h.c.}
\end{eqnarray}
After Fierz rearrangement we obtain the effective SLQ
Lagrangians for the decay $\tau\rightarrow 3\pi\nu_\tau$
\begin{eqnarray}
&&{\cal L}^I_{eff}=-\frac{x_{13}x^{\prime *}_{13}}{2M^2_{\phi_1}}
        \left[(\bar{d}_Lu_R)(\bar{\nu}_{\tau_L}\tau_R)
          +\frac{1}{4}(\bar{d}_L\sigma^{\mu\nu}u_R)
       (\bar{\nu}_{\tau_L}\sigma_{\mu\nu}\tau_R)\right]
          +{\rm h.c.},\nonumber\\
&&{\cal L}^{II}_{eff}=-\frac{y_{13}y^{\prime *}_{13}}{2M^2_{\phi_2}}
        \left[(\bar{d}_Lu_R)(\bar{\tau}^c_R\nu^c_{\tau_L})
         +\frac{1}{4}(\bar{d}_L\sigma^{\mu\nu}u_R)
         (\bar{\tau}^c_R\sigma_{\mu\nu}\nu^c_{\tau_L})\right]\nonumber\\
      &&\hskip 1.5cm +\frac{|y_{13}|^2}{2M^2_{\phi_2}}
        (\bar{d}_L\gamma_\mu u_L)(\bar{\tau}^c_L\gamma^\mu\nu^c_{\tau_L})
        +{\rm h.c.},\nonumber\\
&&{\cal L}^{III}_{eff}=-\frac{|z_{13}|^2}{2M^2_{\phi_3}}
        (\bar{d}_L\gamma_\mu u_L)(\bar{\tau}^c_L\gamma^\mu\nu^c_{\tau_L})
        +{\rm h.c.}
\end{eqnarray}
The tensor part as well as the scalar part in Model I and Model II
can contribute to the $\tau\rightarrow (3\pi)\nu_\tau$ decay.
For simplicity, however, we consider only the scalar contribution in
the present work. Then the size of new contributions from these three
SLQ models is parametrized by the parameters (\ref{decaym})
\begin{eqnarray}
&&\chi^I_{SLQ}=0,\qquad  \hskip 0.3cm
       \eta^I_{SLQ}=-\frac{x_{13}x^{\prime *}_{13}}{2M^2_{\phi_1}},\\
&&\chi^{II}_{SLQ}=-\frac{|y_{13}|^2}{2M^2_{\phi_2}},\qquad
\eta^{II}_{SLQ}=-\frac{y_{13}y^{\prime *}_{13}}{2M^2_{\phi_2}},\\
&&\chi^{III}_{SLQ}=\frac{|z_{13}|^2}{2M^2_{\phi_2}},\qquad
\eta^{III}_{SLQ}=0.
\end{eqnarray}
Note that the vector-type interaction terms have only
real coupling constants, and thus they cannot generate any
CP violating effects. In particular, Model III does not contribute
to CP violation in the decay $\tau\rightarrow 3\pi\nu_\tau$
so that the model will no longer be considered\footnote{\normalsize
\baselineskip 18pt However, in the
semileptonic decays $\tau\rightarrow K\pi\pi\nu_\tau$
and $\tau\rightarrow K\pi K\nu_\tau$, the vector-type interaction
terms from Model II and Model III in principle can have complex couplings
to generate CP-violating effects.}.
In Model I and Model II, the parameters governing CP violation are
\begin{eqnarray}
&&{\rm Im}\xi^I_{SLQ}=-\frac{m^2_{\pi^\prime}}{(m_u+m_d)m_\tau}
 \cdot\frac{{\rm Im}[x_{13}x^{\prime *}_{13}]}{4\sqrt{2}G_FM^2_{\phi_1}},
 \nonumber\\
&&{\rm Im}\xi^{II}_{SLQ}=-\frac{m^2_{\pi^\prime}}{(m_u+m_d)m_\tau}
 \cdot\frac{{\rm Im}[y_{13}y^{\prime *}_{13}]}{4\sqrt{2}G_FM^2_{\phi_2}},
\label{LQp}
\end{eqnarray}
respectively, where all CP-conserving contributions from new physics are
neglected in the normalization. This is justified because the
contributions from new physics are expected to be small compared
to those from the SM.

\subsection{Phenomenological constraints}

\pr
The constraints on the CP-violation parameters,
(\ref{MHDp}) and (\ref{LQp}),
depend upon the values chosen for the $u$ and $d$ quark masses.
In the present work  we use for the light $u$ and $d$ quark
masses\cite{Gasser}
\begin{eqnarray}
m_u=5.0 {\rm MeV},\qquad  \ m_d=9.0 {\rm MeV},
\end{eqnarray}
and for the $W$ boson mass we use $M_W=80$ GeV.
Inserting these values in (\ref{MHDp}) and (\ref{LQp}) we
obtain
\begin{eqnarray}
&&\hskip 0.5cm {\rm Im}(\xi_{MHD})\simeq -1.7\times 10^{-4}
     \frac{[{\rm Im}(XZ^*)-(5/9){\rm Im}(YZ^*)]}{(M_H/M_W)^2},\\
   { }\nonumber\\
&&\left[\begin{array}{c}
      {\rm Im}(\xi^I_{SLQ})\\
      { } \\
      {\rm Im}(\xi^{II}_{SLQ})
      \end{array}\right]
    \simeq
  -7.5\times 10^2
\left[\begin{array}{c}
      \left(\frac{M_W}{M_{\phi_1}}\right)^2
      {\rm Im}(x_{13}x^{\prime *}_{13})\\
      { }\\
      \left(\frac{M_W}{M_{\phi_2}}\right)^2
      {\rm Im}(y_{13}y^{\prime *}_{13})
      \end{array}\right].
\end{eqnarray}
Clearly, sizable CP-violating effects require that ${\rm Im}(XZ^*)$ and
${\rm Im}(YZ^*)$ are large and $M_H$ is small compared to $M_W$ in the
MHD model, and similarly that ${\rm Im}(x_{13}x^{\prime *}_{13})$
and ${\rm Im}(y_{13}y^{\prime *}_{13})$ are large and
$M_{\phi_i}$ ($i=1,2$) are small compared to $M_W$ in the SLQ models.

In the MHD model the strongest constraint\cite{Grossman}
on ${\rm Im}(XZ^*)$ comes from the measurement of the branching ratio
Br($B\rightarrow X\tau\nu_\tau$) which actually gives a constraint
on $|XZ|$. For $M_H<440$ GeV, the bound on ${\rm Im}(XZ^*)$ is given by
\begin{eqnarray}
{\rm Im}(XZ^*)<|XZ|< 0.23 M^2_H {\rm GeV}^{-2}.
\end{eqnarray}
On the other hand, the bound\cite{Grossman} on ${\rm Im}(YZ^*)$
is mainly given by $K^+\rightarrow \pi^+\nu\bar{\nu}$.
The present bound is
\begin{eqnarray}
{\rm Im}(YZ^*)<|YZ|< 110\ \ {\rm for}\ \ m_t=140{\rm GeV}\ \
{\rm and}\ \ M_H=45{\rm GeV}.
\end{eqnarray}
Combining the above bounds and assuming $M_H=45$GeV, we obtain
the bound on ${\rm Im}(\xi_{MHD})$ as
\begin{eqnarray}
|{\rm Im}(\xi_{MHD})|<0.28.
\label{HDc}
\end{eqnarray}

The constraints on the leptoquark couplings can be obtained through
several rare processes\cite{Davidson,Geng,Kane,Hall}.
As a typical bound we find from the experimental
bound  ${\rm Br}(K_L\rightarrow \mu e)<3.3\times 10^{-11}$\cite{PDG94}
for the branching ratio of the lepton-family-number-violating process
$K_L\rightarrow \mu e$,
\begin{eqnarray}
\frac{|x_{21}x^*_{12}|}{M^2_{\phi_1}}< 3\times 10^{-11}{\rm GeV}^{-2},
\label{ModelI}
\end{eqnarray}
and from $\Gamma(\mu {\rm Ti}\rightarrow e {\rm Ti})/
\Gamma(\mu {\rm Ti}\rightarrow
{\rm capture})<4.3\times 10^{-12}$\cite{Dohmen,PDG94},
\begin{eqnarray}
\frac{|y_{11}y^*_{12}|}{M^2_{\phi_2}}< 1.9\times 10^{-11}{\rm GeV}^{-2}.
\label{ModelII}
\end{eqnarray}
On the other hand, the helicity-suppressed $\pi_{e_2}$ decay\cite{Shanker}
can set a strong bound on $|x_{11}x^{\prime *}_{11}|$ and
$|y_{11}y^{\prime *}_{11}|$ under the assumption that
${\rm Im}(x_{11}x^{\prime *}_{11})\sim {\rm Re}(x_{11}x^{\prime *}_{11})$
and
${\rm Im}(y_{11}y^{\prime *}_{11})\sim {\rm Re}(y_{11}y^{\prime *}_{11})$.
Using the experimental value
$R_{exp}=(1.230\pm 0.004)\times 10^{-4}$\cite{PDG94} and
the SM value $R=1.235\times 10^{-4}$\cite{Marciano} for
the ratio of B($\pi\rightarrow e\nu_e(\gamma)$) to
B($\pi\rightarrow \mu\nu_\mu(\gamma)$)
we obtain
\begin{eqnarray}
&&\frac{|{\rm Im}(x_{11}x^{\prime *}_{11})|}{M^2_{\phi_1}}\sim
  \frac{|{\rm Re}(x_{11}x^{\prime *}_{11})|}{M^2_{\phi_1}}
      <7\times 10^{-11}{\rm GeV}^{-2},\nonumber\\
&&\frac{|{\rm Im}(y_{11}y^{\prime *}_{11})|}{M^2_{\phi_2}}\sim
  \frac{|{\rm Re}(y_{11}y^{\prime *}_{11})|}{M^2_{\phi_2}}
      <7\times 10^{-11}{\rm GeV}^{-2}.
\end{eqnarray}
If we assume
${\rm Im}(x_{13}x^{\prime *}_{13})\sim{\rm Im}(x_{21}x_{12}^*)\sim
 {\rm Im}(x_{11}x^{\prime *}_{11})$ and
${\rm Im}(y_{13}y^{\prime *}_{13})\sim{\rm Im}(y_{11}y_{12}^*)\sim
 {\rm Im}(y_{11}y^{\prime *}_{11})$,
then we obtain from (\ref{ModelI}) and (\ref{ModelII}) the following
constraints:
\begin{eqnarray}
|{\rm Im}(\xi^I_{SLQ})|< 1.44\times 10^{-3},\qquad
|{\rm Im}(\xi^{II}_{SLQ})|< 0.90\times 10^{-3}.
\label{LQc}
\end{eqnarray}

Compared to the constraint (\ref{HDc}) on ${\rm Im}(\xi_{MHD})$,
the constraints (\ref{LQc}) on these SLQ CP-violation parameters
are much more severe such that the SLQ models require a larger number
of $\tau$ leptons than are required by the MHD model to detect any
effects of CP violation. However, we should mention here that
the strong constraints (\ref{LQc}) of the leptoquark couplings
are based on the assumption that all the leptoquark Yukawa couplings
are of the same size. This asuumption can fail in a class of
leptoquark models where the lepton-family-number symmetry and
the electron chirality are softly broken.
In such leptoquark models one possible scenario is that
the lepton-family-number-conserving couplings are much larger than
the lepton-family-number-violating couplings, and also
the couplings involving the third lepton generation ($\tau$ and $\nu_\tau$)
are much larger than those involving the first lepton generation
($e$ and $\nu_e$). In such a scenario the constraints (\ref{LQc})
may be too stringent.

One direct constraint on $|x_{13}x^{\prime *}_{13}|$
and $|y_{13}y^{\prime *}_{13}|$ can be provided through the
measurement of B($\tau\rightarrow \pi\nu_\tau$).
We assume that
${\rm Im}(x_{13}x^{\prime *}_{13})\sim {\rm Re}(x_{13}x^{\prime *}_{13})$
and
${\rm Im}(y_{13}y^{\prime *}_{13})\sim {\rm Re}(y_{13}y^{\prime *}_{13})$.
Then, employing the SM value $\Gamma=(2.480\pm 0.025)\times 10^{-13}$
GeV\cite{Marciano} and the experimental value
$\Gamma_{exp}=(2.605\pm 0.093)\times 10^{-13}$
GeV\cite{PDG94} for the $\tau\rightarrow \pi\nu_\tau$ decay width,
we find that
\begin{eqnarray}
&&\frac{|{\rm Im}(x_{13}x^{\prime *}_{13})|}{M^2_{\phi_1}}\sim
  \frac{|{\rm Re}(x_{13}x^{\prime *}_{13})|}{M^2_{\phi_1}}
   < 3\times 10^{-6}{\rm GeV}^{-2},\nonumber\\
&&\frac{|{\rm Im}(y_{13}y^{\prime *}_{13})|}{M^2_{\phi_2}}\sim
  \frac{|{\rm Re}(y_{13}y^{\prime *}_{13})|}{M^2_{\phi_2}}
   < 3\times 10^{-6}{\rm GeV}^{-2}.
\label{direct}
\end{eqnarray}
It is clear that the direct constraints (\ref{direct}) are by far
weaker than the indirect constraints, (\ref{ModelI}) and
(\ref{ModelII}); therefore, the parameters ${\rm Im}(\xi^I_{SLQ})$
and ${\rm Im}(\xi^{II}_{SLQ})$ can in principle be significantly
larger than the upper bounds (\ref{LQc}). We expect
the precise measurement of $\Gamma(\tau\rightarrow\pi\nu_\tau)$
at future $\tau$-Charm factories to give a stronger direct constraint
on $|x_{13}x^{\prime *}_{13}|$ and $|y_{13}y^{\prime *}_{13}|$.

\section{The $\pi^\prime$ decay constant}
\cleqn

\pr
As clearly shown in section~2 every CP-violating observable requires
interference between the $a_1$ mode and the $\pi^\prime$ mode, and
thus the size of CP violation crucially depends on
$f_{\pi^\prime}$ which  determines
the coupling of the $\pi^\prime$ to the weak current. It is, therefore,
important to make a reliable estimate of its magnitude.

The most commonly used value of the $\pi^\prime$ decay constant
in the literature is obtained by a quark model calculation (the `mock'
meson model) by Isgur et al\cite{Isgur}. Especially, the standard $\tau$
decay Monte-Carlo program, TAUOLA\cite{Tauola}, quotes
from Ref.\ \cite{Isgur} the value in the range
\begin{eqnarray}
f_{\pi^\prime}=0.02\sim 0.08{\rm GeV},
\label{eq:fpipTAUOLA}
\end{eqnarray}
where $f_{\pi^\prime}=0.02$ GeV is used as its default value.

However, the decay constant of $\pi'$ is suppressed by the chiral
symmetry which is not taken into account in the mock-meson
calculation, and thus the value quoted in the TAUOLA
might be overestimated.
In this section we reconsider $f_{\pi'}$ in the chiral
Lagrangian framework and show that it actually vanishes
in the chiral limit due to the
mixing\cite{Manohar} of $\pi'$ with the chiral pion field.
$f_{\pi'}$ is then proportional
to the square of the pion mass $m_{\pi}$ and may hence be much smaller
than (\ref{eq:fpipTAUOLA}).

In order to establish notations we start with the usual chiral Lagrangian
with the flavor SU(2)$_{\rm L}\times$SU(2)$_{\rm R}$ symmetry\cite{Bando};
\begin{eqnarray}
  {\cal L}=f^2{\rm tr}(\alpha_{\mu\perp}\alpha^\mu_\perp)
           +\frac{f^2}{4}{\rm tr}(\hat{\chi}+\hat{\chi}^\dagger),
\label{eq:simple_chiral}
\end{eqnarray}
where the matrix-valued $\hat{\chi}$ and the Maurer-Cartan
1-form $\alpha_{\mu\perp}$ are defined as
\begin{eqnarray}
&&\hat{\chi}=2b\xi_L(s+ip)\xi^\dagger_R,\nonumber\\
&&\alpha_{\mu\perp}=-\frac{i}{2}\left[\partial_\mu\xi_L\cdot\xi^\dagger_L
      +i\xi_LL_\mu\xi^\dagger_L-\partial_\mu \xi_R\cdot\xi^\dagger_R
      -i\xi_R R_\mu\xi^\dagger_R\right],
\end{eqnarray}
with the chiral fields $\xi_L$ and $\xi_R$ given by
\begin{eqnarray}
  \xi_R=\xi^\dagger_L={\rm exp}\left[\frac{i\pi^aT^a}{f}\right],
\end{eqnarray}
and $\pi^a$ ($a=1,2,3$) and $T^a$  are the Nambu-Goldstone fields
and the flavor SU(2) group generators, respectively.
The parameter $b$ is given in terms of $u$-quark condensate
$\langle\bar uu\rangle$ and the $\pi$
decay constant $f$ by $b=-\langle\bar{u} u\rangle/f^2 \simeq 1.3$GeV.
The chiral fields $\xi_L$ and $\xi_R$ transform under
SU(2)$_{\rm L}\times$SU(2)$_{\rm R}$ as
\begin{eqnarray}
  \xi_L \rightarrow \xi_L' = h \xi_L g_L^\dagger, \qquad
  \xi_R \rightarrow \xi_R' = h \xi_R g_R^\dagger,
\end{eqnarray}
where the SU(2) group element $h$ is introduced to maintain the relation
$\xi_R'=\xi_L^{\prime\dagger}$.
In the Lagrangian (\ref{eq:simple_chiral}) we have also introduced
external scalar and pseudoscalar fields, $s$ and $p$, and external
left- and right-handed flavor gauge fields, $L_\mu$ and
$R_\mu$\cite{Gasser2}.
In the real world the external scalar and pseudoscalar
fields are fixed to be $s=\mbox{diag}(m_u, m_d)$ and $p=0$.

We next describe how to introduce a massive pseudoscalar $\pi^\prime$
in the chiral Lagrangian formalism to discuss its physical properties.
Since  $\pi^\prime$  is not a Nambu-Goldstone boson it should be
treated as a matter field transforming as
\begin{eqnarray}
P\rightarrow hPh^\dagger,
\end{eqnarray}
where $P$ is the $\pi'$ field.
The low-energy effective Lagrangian of $\pi$ and $\pi'$ with the lowest
derivatives can be written as
\begin{eqnarray}
  {\cal L}_P
  = {\cal L}_{\rm chiral}
    + \mbox{tr}(D_\mu P D^\mu P) - M_P^2 \mbox{tr}(P^2)
    + i G_P \mbox{tr}((\hat \chi - \hat \chi^\dagger) P)
    + 2 F_P \mbox{tr}(\alpha_{\mu\perp} D^\mu P),
\end{eqnarray}
where $F_P$ and $G_P$ are parameters describing the low-energy
properties of the $\pi'$.
The covariant derivative of the field $P$ is given by
\begin{eqnarray}
  D_\mu P = \partial_\mu P - i [\alpha_{\mu\parallel}, P],
\end{eqnarray}
with the Maurer-Cartan 1-form
\begin{eqnarray}
\alpha_{\mu\parallel}=-\frac{i}{2}\left[\partial_\mu\xi_L\cdot\xi^\dagger_L
      +i\xi_LL_\mu\xi^\dagger_L+\partial_\mu \xi_R\cdot\xi^\dagger_R
      +i\xi_R R_\mu\xi^\dagger_R\right].
\end{eqnarray}

First, we discuss the chiral limit where $\hat\chi=0$. In this limit
the Lagrangian ${\cal L}_P$ is expanded as
\begin{eqnarray}
&&{\cal L}_{P}
  =\frac{1}{2}(\partial_\mu \pi^a)^2
    +\frac{1}{2}(\partial_\mu P^a)^2\nonumber\\
&&\hskip 0.8cm -\frac{F_P}{f} (\partial_\mu \pi^a)(\partial^\mu P^a)
    -\frac{M_P^2}{2} (P_a)^2
    + f \partial_\mu \pi^a  A^{a\mu}
    - F_P \partial_\mu P^a  A^{a\mu}
    + \cdots,
\label{eq:expand_component}
\end{eqnarray}
where $A^a_\mu$ is an external axial vector field,
$A^a_\mu=(R^a_\mu-L^a_\mu)/2$.
Although the nonvanishing $F_P$ in the last term of
(\ref{eq:expand_component}) seems to directly indicate the
existence of the $\pi'$ decay constant
the existence of $F_P$ simultaneously causes a kinetic mixing
of $\pi$ and $P$ through the third term of
(\ref{eq:expand_component}), and thus we need to diagonalize the
kinetic mixing so as to evaluate the actual size of the $\pi'$ decay
constant.

We resolve this kinetic mixing by the redefinition of fields:
\begin{eqnarray}
  \pi_0^a = \pi^a - \left(\frac{F_P}{f}\right) P^a, \qquad
  P_0^a   = \sqrt{1-\left(\frac{F_P}{f}\right)^2} P^a.
\label{Redefinition}
\end{eqnarray}
We find that the field redefinition pushes the $\pi^\prime$
mass to a higher value,
\begin{eqnarray}
m_{\pi^\prime}^2 = \frac{\displaystyle M_P^2}
                    {\displaystyle{1-\left(\frac{F_P}{f}\right)^2}},
\end{eqnarray}
and it forces the $\pi^\prime$ decay constant to vanish in the chiral limit:
\begin{eqnarray}
  f_{\pi^\prime}=0.
\label{eq:fpipchiral}
\end{eqnarray}
Although our calculation is based on the effective Lagrangian with the
lowest derivatives, the result (\ref{eq:fpipchiral}) itself is rather
general.
Actually, we can show that the existence of the higher
derivative terms do not change this result.

In the real world $u$ and $d$ masses are not zero and break the
chiral symmetry explicitly.
Including the $\pi$-mass term due to the finite $u$ and $d$ quark masses
and using the previous field redefinition (\ref{Redefinition})
we obtain the mass term of the $\pi$ and $\pi^\prime$ chiral Lagrangian as
\begin{eqnarray}
{\cal L}_M
   \! =\! -\frac{\hat{m}_{\pi}^2}{2}(\pi_0^a)^2
     \! -\!\hat{m}_{\pi}^2\frac{F_P+2G_P}{f\sqrt{1-F_P^2/f^2}}
       \pi_0^a P_0^a
     \! -\!\frac{\hat{m}_{\pi}^2}{2(1-F_P^2/f^2)}
       \left(\frac{M_P^2}{\hat{m}_{\pi}^2}+4\frac{G_P F_P}{f^2}\right)
       (P_0^a)^2,
\label{eq:mass}
\end{eqnarray}
with $\hat m_\pi^2 = (m_u+m_d)b$.
We diagonalize the Lagrangian (\ref{eq:mass}) in the perturbation
of $\hat{m}_{\pi}^2$:
\begin{eqnarray}
\pi_0^a &=& \pi_{\rm phys}^a-\hat{m}_{\pi}^2\cdot\delta\cdot P_{\rm phys}^a
  +\cdots,
  \nonumber\\
P_0^a
  &=& P_{\rm phys}^a + \hat{m}_{\pi}^2 \cdot\delta\cdot\pi_{\rm phys}^a
  +\cdots.
\label{eq:diag}
\end{eqnarray}
Plugging (\ref{eq:diag}) into the Lagrangian (\ref{eq:mass})
we find that the parameter $\delta$ removing the mixing term is
\begin{eqnarray}
  \delta = - \sqrt{\displaystyle 1 - \frac{F_P^2}{f^2}}
               \cdot\left(\frac{F_P+2G_P}{M_P^2 f}\right).
\end{eqnarray}
The complete diagonalization leads to
\begin{eqnarray}
m_\pi^2 = \hat{m}_{\pi}^2, \qquad
m_{\pi^\prime}^2=\frac{M_P^2+4\hat{m}_{\pi}^2 G_PF_P/f^2}{1-F_P^2/f^2}
             \simeq \frac{M_P^2}{1-F_P^2/f^2}.
\end{eqnarray}
On the other hand, the meson coupling term to the axial vector current
$A^a_\mu$ are written in terms of the physical fields $\pi^a_{\rm phys}$
and $P^a_{\rm phys}$ as
\begin{eqnarray}
f A_\mu^a \partial^\mu \pi_0^a
  = f A_\mu^a \partial^\mu \pi_{\rm phys}^a
     +\frac{\hat{m}_{\pi}^2}{M_P^2}\sqrt{\displaystyle 1 - F_P^2/f^2}
      (F_P+2G_P)A_\mu^a \partial^\mu P_{\rm phys}^a,
\label{current}
\end{eqnarray}
and then the $\pi$ and $\pi^\prime$ decay constants directly read
\begin{eqnarray}
  f_\pi = f, \qquad
  f_{\pi^\prime}=\frac{m_\pi^2}{M_P^2}\sqrt{\displaystyle 1 - F_P^2/f^2}
                 (F_P+2G_P).
\end{eqnarray}
As a result, we obtain the following expression:
\begin{eqnarray}
  r_\pi \equiv \frac{f_{\pi'}^2 m_{\pi'}^4}{f_\pi^2 m_\pi^4}
        =  \frac{F_P^2}{f_\pi^2 (1-F_P^2/f_\pi^2)}
           \left(1 + 2\frac{G_P}{F_P}\right)^2.
\label{eq:rpi}
\end{eqnarray}
It is now clear that $f_{\pi^\prime}$ is
proportional to the square of the pion mass $m_\pi$, and thus it
is naturally expected that the value of $f_{\pi^\prime}$ is small.

The parameters $F_P$ and $G_P$ are expected to be of the size of QCD
scale.
We thus naively expect
\begin{eqnarray}
  r_\pi = {\cal O}(1).
\label{eq:naive}
\end{eqnarray}
On the other hand, the value of the $f_{\pi'}$ quoted in TAUOLA
(\ref{eq:fpipTAUOLA}) leads to a large value of $r_\pi$,
\begin{eqnarray}
  r_\pi = 300 \sim 6000,
\end{eqnarray}
which clearly contradicts our naive expectation (\ref{eq:naive}).

Although the precise values of those parameters cannot be determined
within the chiral Lagrangian framework, we can estimate them from the
QCD dynamics.
In the present work, we employ the non-relativistic chiral
quark (NRCQ) model of Manohar and Georgi\cite{Manohar} and
QCD sum rules\cite{Narison} to make a rough estimate of the parameters.

In the NRCQ model the lowest S-wave bound states of the chiral quarks are
identified as $\rho$ and $\pi^\prime$.
Neglecting the spin-dependent interaction we can take
\begin{eqnarray}
  M_P \simeq m_\rho = 0.773\mbox{GeV}.
\end{eqnarray}
Both $G_P$ and $F_P$ are calculated from the wave function at the
origin.  In the non-relativistic approximation we find the relation
\begin{eqnarray}
  2b G_P \simeq M_P F_P
\end{eqnarray}
On the other hand the value of $F_P$ can be estimated from the relation
\begin{eqnarray}
  m_{\pi^\prime}^2 \simeq \frac{M_P^2}{1-F_P^2/f^2_{\pi}}
   = (1.3\mbox{GeV})^2,
\end{eqnarray}
with $f_\pi=0.0933$ GeV.
It is now straightforward to show that
\begin{eqnarray}
  r_\pi \simeq 4,\qquad
  f_{\pi^\prime}\simeq 2\times 10^{-3}{\rm GeV}.
\end{eqnarray}

Let us next discuss the evaluation of $r_\pi$ with the QCD sum rule
technique\cite{Narison}.
By using the operator product expansion of the isotriplet scalar
current, $j_a^{(a_0)}$, and
the pseudoscalar current, $j^{(\pi)}_a$, we find a sum rule,
\begin{eqnarray}
  \int_0^\infty ds {\rm e}^{-s/M^2} {\rm Im} \Pi(s) \simeq 0,
\label{eq:sumrule}
\end{eqnarray}
where $\Pi(s)$ is defined by
\begin{eqnarray}
  \Pi(-q^2) \delta^{ab} \equiv i \int dx {\rm e}^{iqx}
            \langle T\{j^{(\pi)}_a(x)j^{(\pi)}_b(0)
                     - j^{(a_0)}_a(x)j^{(a_0)}_b(0)\} \rangle,
\end{eqnarray}
and $M$ is the scale of the Borel transform.
Assuming the $a_0$ (isotriplet scalar meson), $\pi$ and $\pi'$ meson
dominance we can derive an expression\cite{Narison} for $r_\pi$;
\begin{eqnarray}
r_\pi=-\left(\frac{m^2_{a_0}-m^2_\pi}{m^2_{a_0}-m^2_{\pi^\prime}}\right)
      {\rm exp}[-(m^2_\pi-m^2_{\pi^\prime})/M^2],
\end{eqnarray}
 from (\ref{eq:sumrule}) and its first derivative of $1/M^2$.
Inserting the $\pi$ and $\pi^\prime$ masses and $m_{a_0}=0.98$ GeV, we find
that
\begin{eqnarray}
r_\pi\simeq 7,\qquad
f_{\pi^\prime}\simeq 3\times 10^{-3}{\rm GeV},
\end{eqnarray}
for the mass scale $M$ around 1 GeV.

Note that despite their very different approaches,
the NRCQ model\cite{Manohar} and the QCD sum rule\cite{Narison}
give a similar estimate of the $r_\pi$. According to those estimates
one find $f_{\pi^\prime}\approx (2\sim 3)\times 10^{-3}$ GeV,
which is much smaller than the value ($0.02$ GeV) quoted in TAUOLA.
Considering possible uncertainties
in our estimates we use in our numerical
analysis a slightly broader range of $f_{\pi^\prime}$:
\begin{eqnarray}
f_{\pi^\prime}=(1\sim 5)\times 10^{-3}{\rm GeV}.
\label{fpi_range}
\end{eqnarray}
Certainly a more exact value of $f_{\pi^\prime}$ can be determined directly
from experiment.

\section{Numerical results}
\cleqn

\pr
For the numerical analysis we use for the CP-violation parameter values
in the MHD model and the SLQ models:
\begin{eqnarray}
&&{\rm Im}(\xi_{MHD})=0.28,\\
&&{\rm Im}(\xi^I_{SLQ})=1.44\times 10^{-3},\qquad \
  {\rm Im}(\xi^{II}_{SLQ})=0.90\times 10^{-3},
\label{Bound}
\end{eqnarray}
which are maximally allowed values according to the constraints (\ref{HDc})
and (\ref{LQc}). We mention in passing once more that the
constraints (\ref{LQc}) on the leptoquark couplings may be
too stringent in a class of leptoquark models where the
leptoquark Yukawa couplings are significantly generation-dependent.
For the $\pi^\prime$ decay constant we consider the range:
$(1\sim 5)\times 10^{-3}$ GeV as in eq.~(\ref{fpi_range}).

Fig.~2(a) shows the normalized differential decay width,
$\Sigma^{-1}[{\rm d}\Sigma/{\rm d}\sqrt{q^2}]$, for the decay
$\tau\rightarrow (3\pi)\nu_\tau$ as a function of the $(3\pi)$
invariant mass, $\sqrt{q^2}$. Note that it has almost one peak.
The reason is not only because $m_{\pi^\prime}$ and
$m_{a_1}$ are quite similar in magnitude, but also because the $\pi^\prime$
contribution is by far smaller than the $a_1$ contribution. This
justifies the approximation where all CP-conserving contributions
from new scalar-fermion interactions are neglected in the normalization.
In Fig.~2(b) we present the two CP-violating differential forward-backward
asymmetries, ${\rm d}A_{1FB}/{\rm d}\sqrt{q^2}$
and ${\rm d}A_{2FB}/{\rm d}\sqrt{q^2}$, as a function
of the $(3\pi)$ invariant mass $\sqrt{q^2}$. In this case we take
$f_{\pi^\prime}=5\times 10^{-3}$ GeV
and ${\rm Im}(\xi)=0.28$.
The two asymmetries have a similar dependence on $\sqrt{q^2}$, but
it is clear that the asymmetry $A_{1FB}$ is almost 4 times larger
than the asymmetry $A_{2FB}$.

Table~1 shows the expected size of the integrated asymmetries
$A_{1FB}$ and $A_{2FB}$ and the optimal asymmetry
$\varepsilon_{opt}$  along with the number of $\tau$
leptons, $N^\tau$, required to obtain the $2\sigma$ signal in the MHD model
and the two SLQ models for the CP-violation parameter
values (\ref{Bound}) and for the $\pi^\prime$ decay constant in the
range (\ref{fpi_range}).
The optimal asymmetry, $\varepsilon_{opt}$, is optimized with
respect to the kinematic variables ($q^2,s_1,s_2,\cos\theta,\phi^*)$
so that its expected size denotes the maximally obtainable
asymmetry in the three-pion decay mode.
However, we find that the CP-violating forward-backward asymmetry
$A_{1FB}$ is sizable, while the other asymmetry,
$A_{2FB}$, is rather small.
The $A_{1FB}$ asymmetry and the corresponding values of $N^\tau$ given
in Table~1 show that the CP-violating effects in the MHD model may be seen
with less than $10^8$ $\tau$'s even for a rather small
$\pi^\prime$ decay constant.
To see the CP-violating effects with the asymmetries $A_{1FB}$
and $A_{2FB}$ in the SLQ models more than
$10^{11}$ $\tau$'s may be required.
On the other hand, the $\varepsilon_{opt}$ and the corresponding
$N^\tau$ values show that the CP-violating effects from the MHD model
can be detected with about $10^7$ $\tau$'s, while still more than
$10^{10}$ $\tau$ leptons are required to see CP violation in the SLQ models.
However, as previously mentioned, the CP-violating parameters
${\rm Im}(\xi^I_{SLQ})$ and ${\rm Im}(\xi^{II}_{SLQ})$ may be considerably
larger so that the required number of $\tau$ leptons could be significantly
reduced. In light of the fact that about $10^7$ and $10^8$
$\tau$ leptons are produced yearly at $B$ factories and
$\tau$-Charm factories, respectively, we conclude that CP violating
effects for the MHD model may be seen at the $\tau$-Charm factories
through the CP-violating forward-backward asymmetry, $A_{1FB}$, and
even at the $B$ factories with the optimal asymmetry, $\varepsilon_{opt}$,
for the $\pi^\prime$ decay constant in the range (\ref{fpi_range}).
Finally, we note that, if we take $f_{\pi^\prime}=0.02$GeV, the value quoted
in TAUOLA , the number of $\tau$ leptons needed is reduced by a factor
of at least 15.

\section{Conclusion}

\pr
In this paper, we have investigated the possibility of probing CP violation
through the semileptonic decays
$\tau\rightarrow (3\pi)\nu_\tau$ where two resonances,
$a_1$ and $\pi^\prime$, contribute.
The interference of the $a_1$ and $\pi^\prime$ resonances
leads to  enhanced CP-violating asymmetries whose magnitude
crucially depends on the $\pi^\prime$ decay constant $f_{\pi^\prime}$.
We made an estimate of $f_{\pi^\prime}$ with a simplified chiral
Lagrangian coupled to a massive pseudoscalar field, and we
compared the estimates from the nonrelativistic chiral quark model
and from the QCD sum rule with the estimate from the `mock' meson model.
We found that the mixing between the pion field and the massive
pseudoscalar field renders $f_{\pi^\prime}$ proportional to
the square of the pion mass, $m_\pi$, and hence it is much
smaller than the value quoted in the $\tau$ library, TAUOLA.
We considered two CP-violating forward-backward asymmetries,
$A_{1FB}$ and $A_{2FB}$, and the optimal asymmetry, $\varepsilon_{opt}$,
which is optimized with all the five kinematic variables
$(q^2,s_1,s_2,\cos\theta, \phi^*)$. With these asymmetries we
quantitatively estimated the size of CP violating effects from the
multi-Higgs-doublet model and the scalar-leptoquark models.
We found that while the scalar-leptoquark models may require
more than $10^{10}$ $\tau$ leptons for the detection of any effect of CP
violation, CP-violating effects from the multi-Higgs-doublet model
may be seen at the $2\sigma$ level with about $10^7$ $\tau$ leptons
using the chiral Lagrangian
estimate of $f_{\pi^\prime}=(1\sim 5)\times 10^{-3}$ GeV.

\newpage

\section*{Acknowledgements}

S.Y.~Choi would like to thank the Japanese Ministry of Education,
Science and Culture (MESC) for the award of a visiting fellowship.
The authors would like to thank R.~Szalapski for carefully reading
the paper. This work was supported in part by the Grant-in-Aid
for Scientific Research from MESC (No. 05228104).

\newpage

\section*{References}

\newpage

\section*{Tables}
\renewcommand{\labelenumi}{\bf Table {\arabic {enumi}} \\}

\begin{enumerate}

\vspace*{0.3cm}
\item{
The expected size of the CP-violating forward-backward asymmetries,
$A_{iFB}$ ($i=1,2$), and the optimal asymmetry,
$\varepsilon_{opt}$, and the number of $\tau$ leptons, $N^\tau$,
needed for detection at the $2\sigma$ level are determined
for the range $f_{\pi^\prime}=(1\sim 5)\times 10^{-3}$GeV
with ${\rm Im}(\xi_{MHD})=0.28$ in the MHD model, and
${\rm Im}(\xi^I_{SLQ})=1.44\times 10^{-3}$ and
${\rm Im}(\xi^{II}_{SLQ})=0.90\times 10^{-3}$
in the two SLQ models.}

\vspace*{1.2cm}

\begin{tabular}{|c|c|c|c|c|}\hline
{ } & { } & { } & { } &{ } \\
 Asymmetry  & Size(\%) &
            $N_{MHD}^\tau(10^7)$  &
            $N_{SLQI}^\tau(10^{11})$ &
            $N_{SLQII}^\tau(10^{11})$\\
{ } & { } & { } & { } & { }\\
\hline
{ } & { } & { } & { } & { } \\
$A_{1FB}$ & $0.05\sim 0.25$  &
       $ 12\sim 300$ &
       $ 45\sim 1125$ &
       $ 120\sim 3000$\\
{ } & { } & { } & { } & { }\\
$A_{2FB}$ & $0.20\sim 1.00$ &
       $ 0.75\sim 19$ &
       $ 2.8\sim 72$ &
       $ 7.4\sim 185$\\
{ } & { } & { } & { } & { }\\
$\varepsilon_{opt}$& $0.48\sim 2.4$  &
       $ 0.13\sim 3.3$ &
       $ 0.49\sim 12$ &
       $ 1.3\sim 33$\\
{ } & { } & { } & { } & { }\\
\hline
\end{tabular}
\end{enumerate}

\newpage

\section*{Figures}
\vskip 1cm
\begin{enumerate}

\item[{\bf Figure 1}] The decay $\tau^-\rightarrow (3\pi)^-\nu_\tau$
     viewed from the (a) $\tau^-$ and (b) $(3\pi)^-$ rest frames.
     The two frames are related by a Lorentz boost along the tau neutrino
     direction and through a rotation by $\theta$ with respect to
     the common $y$-axis.
\vskip 0.5cm
\item[{\bf Figure 2}]
     (a) A plot of the normalized differential decay width,
     $\Sigma^{-1}[{\rm d}\Sigma/{\rm d}\sqrt{q^2}]$,
     as a function of the $(3\pi)$ invariant mass, $\sqrt{q^2}$.
     (b) CP-violating asymmetries as a function of the $(3\pi)$
     invariant mass, $\sqrt{q^2}$, with the $\pi^\prime$
     decay constant, $f_{\pi^\prime}=5\times 10^{-3}$GeV
     as a reference value.
     The solid line is for the CP-violating
     forward-backward asymmetry, ${\rm d}A_{1FB}/{\rm d}\sqrt{q^2}$,
     and the long-dashed line is for the CP-violating forward-backward
     asymmetry, ${\rm d}\hat{A}_{2FB}/{\rm d}\sqrt{q^2}$.

\end{enumerate}

\end{document}